\documentclass[11pt,twoside]{article}
\usepackage{csdyn}

\resetcounters

\begin{document}

\title{Circumstellar disks during various evolutionary stages}
\author{Ren\'e D. Oudmaijer 
\affil{School of Physics and Astronomy \\ University of Leeds \\ Leeds  LS2
  9JT, UK}}

\begin{abstract}
Disks are ubiquitous in stellar astronomy, and play a crucial role in
the formation and evolution of stars.  In this contribution we present
an overview of the most recent results, with emphasis on high spatial
and spectral resolution. We will start with a general discussion on
direct versus indirect detection of disks, and then traverse the HR
diagram starting with the pre-Main Sequence and ending with evolved
stars.
\end{abstract}

\section{Introduction}

Disks and disk-like structures are ubiquitous in astrophysics, they
can be found at small scales as rings around planets in our Solar
System, which in turn were formed in a disk surrounding the young
Sun. At much larger scales we find the accretion disks around the
central Black Holes that power Active Galactic Nuclei, while larger
still, we know of course of disky galaxies.  From a stellar evolution
point of view, they provide crucial information about the formation of
stars, as these grow via the accretion through a disk. At later stages
the disk-like structures are associated with the shaping of the ejecta
of evolved stars, such as the beautiful Planetary Nebulae and probably
also the rings seen around SN1987A.

The {\it direct} detection of circumstellar disks has always been a
challenge. Particularly so for the inner parts where all the action
happens. Not surprisingly, this is because most of the emission comes
from the inner disks. To give an idea of the scales involved, for
example the bulk of the ionized gas that gives rise to hydrogen
recombination lines is often confined to distances of order several
stellar radii from the star. Up until the turn of the century, we
mostly had to rely on indirect techniques to study this material, and
it is only very recently that we have been able to spatially resolve
such disks on a more-or-less routine basis.

Historically, spectroscopy has been a powerful diagnostic for the
presence of disks.  As far back as the thirties, in a seminal paper
\citet{struve_1931} not only hypothesized that the doubly peaked
H$\alpha$ recombination emission lines observed towards bright Be
stars originated from a rotating disk, he also backed this up with
model simulations. This comparison of data with models is a very early
application of an approach of which we will see many examples during
this workshop.  Although very compelling, Struve's disk hypothesis
suffered from the fact that the spectroscopy can not provide {\it
direct} evidence for a disk. The quest to prove once and for all that
Be stars are surrounded by disks has taken several decades. We should
not omit the important contribution of spectropolarimetry in this
quest. As explained in much more detail by Magalh{\~a}es in these
proceedings, spectropolarimetry is a very powerful method to probe
circumstellar material close to a star. It takes advantage of the fact
that due to collisions with electrons in an ionized medium, the light
from the star is much more polarized than the emission lines such as
H$\alpha$ which originate from the gas itself. As only non-round
geometries will result in a net polarization, the difference in
polarization levels over H$\alpha$ and the continuum serve as
excellent diagnostics of very small scale disks.
\citet{poeckert_1977} refined this technique and
computed the polarization signatures from rotating disks. In a series
of papers, they applied this method to Be stars and demonstrated that
disks are the best explanation for the observations.  The final
chapter in this story is the interferometric image of the Be star
$\zeta$ Tau, taken in a narrow band H$\alpha$ filter presented by
\citet{quirrenbach_1994}. The elongated structure clearly reveals the
presence of a disk, with a position angle on the sky consistent with
the polarimetry.  This paved the way for further detailed studies and it
has now even become possible to both spatially and spectrally probe
the disks and their kinematics. For example, very recently it could be
determined that the disks rotate Keplerian \citep{wheelwright_2012mn,
kraus_2012}, using spectro-astrometry and VLTI interferometry
respectively. This finding allows us to move forward and investigate
the disk origins with much better observational constraints. I will
refrain from reporting on more recent work on Be stars as these are
covered in many other contributions in these proceedings.

This historical note also tells us the importance of Be stars in the
grander scheme. Due to their brightness and proximity, many new
observational techniques, including those mentioned above, are tried
and tested first on the bright and nearby Be stars.  Later, when the
detection techniques have improved,  often in combination with
larger telescopes, the methods can then be applied to other types of
object.  It should also be noted that the direct confirmation of a
disk by interferometry validates indirect methods such as the
spectroscopy and polarimetry as viable disk diagnostics. We should of
course keep in mind that there still remain caveats that are
inevitably associated with indirect methods. Spectroscopy for example
can often be interpreted in many ways, while even rotating disks do
not necessarily display doubly peaked lines
(e.g. \citealt{elitzur_2012}).

With the exception of the Be stars, it is fair to say that at the turn of this
century, most evidence and information derived for circumstellar disks
concerned indirect methods.  Since then, we have seen a wealth of new methods
to detect and study disks in detail.  These include spectro-astrometry
(e.g. Wheelwright, these proceedings), optical/NIR interferometry (Stee, Groh,
Millour all in these proceedings) and integral field spectroscopy (Stecklum,
these proceedings).  Last but not least, a significant development is the
increased interplay with highly sophisticated models to interpret these data
(Bjorkman, Carciofi, Jones, these proceedings).

In the following, I highlight several examples of disks and disk-like
structures around stars at various stages of their evolution.  I will
concentrate on high resolution studies in the top of the HR diagram,
which contains both young and evolved objects. I tried to avoid too
much overlap with other presentations at this meeting. Even so, given the
space constraints, it will be a necessarily shallow tour, which by no
means can be regarded as complete, and much good work may not be
cited.  The citations that are given should serve as a good start in
any search for relevant references. Also, the review may seem
optically/NIR biased, this is simply because at present the inner
parts of the disks are most efficiently traced by diagnostics at the
shorter wavelengths.

I should finish this introduction with a cautionary remark. Any
high-resolution observation, be it spatial, spectral or temporal, is
very time consuming. The papers describing the state-of-the-art are
therefore often necessarily single-object papers. It is tempting to
extrapolate the conclusions reached for individual objects to the
general class.  Instead, this is the best opportunity to use these
detailed studies to support the indirect diagnostics for which much larger
samples are available so that more robust general conclusions can be drawn.

\section{Where it starts, the pre-Main Sequence}

The ambiguity associated with indirect methods is well illustrated by
the lively debate in the nineties about the nature of the
circumstellar material around the intermediate mass pre-Main Sequence
Herbig Ae/Be stars. The debate concerned whether Herbig Ae/Be stars
were surrounded by disks or not (see e.g. \citealt{waters_1998} for an
overview at that time). The question was relevant as the formation of
massive stars was, and in many ways still is, a matter of uncertainty.
A key difference with the formation of lower mass, Sun-like stars, is
that the lower mass stars sustain magnetic fields, and grow via
magnetically controlled accretion through a disk. On the other hand,
higher mass stars with radiative envelopes are not expected to form in
this manner.  To make matters worse, a spherical accretion geometry
has its own difficulties as strong radiation pressure from these
hotter stars may halt accretion altogether.  Recently
\citet{krumholz_2009} were able to show from high resolution
computations that massive stars can indeed form by disk accretion. The
outflows are channeled through the bipolar cavity and consequently the
disk undergoes less radiation pressure. A side result of their
calculations is that due to fragmentation of the original disk
structure, massive stars are predominately formed in roughly equal
mass binaries at separations of order au. A prediction which we come
back to later.

Observationally, the study of massive young stars is, amongst others, hampered
by the fact that they are much rarer than lower mass stars, and therefore much
further away, requiring high resolution studies to probe the accretion
region. However, imaging of these objects was not possible for a long time,
and astronomers resorted to the next best diagnostic, the Spectral Energy
Distribution (SED).  The trouble however is that fits of SEDs are
degenerate. They can be readily explained by both disks and spherical
envelopes, with either possibilities having had their own proponents (e.g
\citealt{pezzuto_1997}). Headway was only made when \citet{mannings_1997}
detected rotating disk-like structures around some Herbig stars, and the
presence of disks was seemingly established. Not much later, \citet{miro_1999}
unified both SED scenarios by demonstrating that the simultaneous presence of
both components can explain the SED; the inner, warm, disk contributes to the
NIR excess, while a cooler spherical envelope dominates the longer wavelength
excess emission. The case of disk accretion for Herbig Ae/Be was not yet
settled however, as the larger (200-600 au) disk-like structures observed at
mm wavelengths are too far from the star to probe accretion. Interestingly,
the issue that many such disk tracers only probe regions far out and thus can
not be used to study the accretion process is often overlooked.

\articlefigure{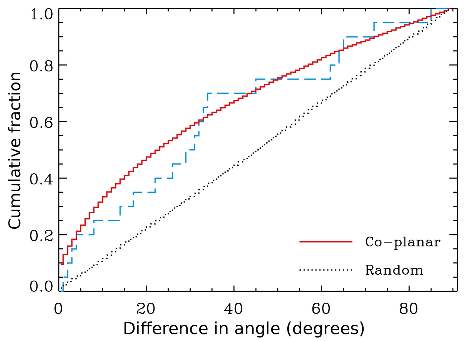}{lab_fig} { This figure shows the cumulative
  fraction of the difference between the disk polarization angles and
  the binary PAs of a sample of Herbig stars (blue dashed). When the
  angle is 90$^{\rm o}$, the orbit and disk have the same position
  angle on the sky. If both are aligned (i.e. co-planar), a
  distribution as presented by the red solid line would be observed
  (due to, amongst others, inclination effects, not every co-planar
  system would have a disk and orbit with the same observed PA).  The
  black dotted line represents the distribution if the two angles were
  randomly distributed. The co-planar scenario is the most likely
  situation, lending support to the disk-fragmentation scenario by
  Krumholz and collaborators \citep[see text]{wheelwright_2011}.}

The most significant progress in this field over the last years is
that optical/infrared interferometric instrumentation made
observations of the inner disks at milli-arcsecond scales
possible. This is at scales orders of magnitude smaller than the outer
disks detected by infrared and (sub-)mm observations.  The inner disks
were first observed using single baselines which allowed basic size
measurements. It was found that the inner boundaries of the disks get
larger as a function of luminosity, consistent with the notion that
the dust sublimation radius is measured (as documented by
e.g. \citealt{millangabet_2007}). However, not all Herbig Ae/Be stars
follow this relationship, and some objects turned out to be smaller
than predicted. This has led to several suggestions for the nature of
the near-infrared emission. Competing ideas include that the
near-infrared emission comes from optically thick gas
(\citealt{kraus_2008}), or that it is due to refractory grains which
can survive higher temperatures (\citealt{benisty_2010}).  Full
model-independent images of several Herbig Ae/Be stars have now been
published, proving beyond doubt that the objects are surrounded by
disks (e.g. \citealt{benisty_2011} and references therein).  The
number of imaged objects is still limited. Improved sensitivity and
instruments that can combine more baselines such as those on CHARA,
MROI and VLTI are either just installed and planned respectively,
making it much easier to obtain images and paving the way for
statistical studies. In parallel, the next step is to fully understand
and parametrize the properties of the disks in order to follow the
formation and evolution of intermediate and massive stars.  An early
example has been published by \citet{weigelt_2011} who took advantage
of the high spectral resolving power of VLTI/AMBER ($R \sim 12000$)
and spectrally resolved the Br$\gamma$ line of the Herbig Be star MWC
297. By fitting the spectral line, they find that the emission is due
to a disk-wind.

Furthermore, returning to the point originally made about sample
sizes, informed by the fact that the Herbig Ae/Be stars are surrounded
by disks, \citet{wheelwright_2011} use their large sample of objects
with spectropolarimetric data which trace the disks and apply it to
the (many) known binaries in the class.  They find that the disks
around the primary objects are aligned with the orbital planes of the
binaries (see Figure 1).  Most Herbig Ae/Be stars are found in binary
systems with separations of order au's, while their mass ratios peak at
unity \citep{hugh_2010}. With the new finding that the orbital planes
and the disks are aligned, the prediction by \citet{krumholz_2009}
that the disk fragmentation leads to binaries with the same properties
as now observed, is confirmed.

Finally, there are the more embedded Massive Young Stellar Objects
with masses exceeding 10-15M$_{\odot}$, these are on average even
further away than Herbig Ae/Be stars, with as added complication that
they are optically invisible due to the large extinction.  De Wit
\citep[these proceedings;][]{dewit_2011} presents tantalizing evidence
for disks around these most massive young stars as well.

\section{Moving on, to the Main Sequence}

The disks concerned in the previous section are the accretion disks
which formed as a result from the collapse of a rotating cloud. After
the accretion has halted, and the star is settling onto the Main
Sequence, the disks gradually disappear, mostly due to
photo-evaporation.  During this phase, the disks are commonly referred
to as proto-planetary or transition disks. Later, after most dusty
particles should have been lost due to processes such as the
Poynting-Robertson effect and radiation pressure, the remaining
``debris'' disk's dust reservoir is replenished by collisions of larger
bodies.  Before the disks have disappeared altogether, planets must
have formed.

The transition disks are the birth places and thus important
laboratories to study the formation of planets. Evidence that planets
are indeed present in these disks has emerged over the past decade.
Early SEDs of Herbig Ae/Be stars and their evolutionary successors
were found to display both a hot and a cool component suggesting the
presence of a single disk with a gap in which the dust had been
cleared (\citealt{malfait_1998}).  These gaps were confirmed by high
resolution observations at sub-mm and mm wavelengths (for an overview
see the paper by \citealt{williams_2011}), however they are located
relatively far from the star by virtue of the sub-arcsecond resolution
of the data.  The inner au's were probed by spectro-astrometry
\citep{pontoppidan_2008}, while optical interferometry revealed a hole
in the disk around HD 100546 at a distance of less than 10 au from the
star \citep{tatulli_2011}.

Such gaps are now widely accepted as being due to a giant planet orbiting the
star. This was basically confirmed since planets have been found around
objects with disks. For example, $\beta$ Pic and Fomalhaut both have a
proto-planetary disk and host a planet (\citealt{lagrange_2010, kalas_2008}),
while T Cha has recently been reported to have a planet located in the gap
itself (\citealt{krausa_2012}). Despite these successes, it is very hard to
detect planets around active objects surrounded by circumstellar material. The
holes in the disks as indicated by the SEDs can be used as a proxy for the
existence of planets around these stars. The ensuing statistics may tell us a
great deal about the timescales involved in the formation of planets.
Ironically, a current line of research investigates whether the planets
themselves play a dominant role in the disk dispersal.

Other disks around Main Sequence stars are known, such as those around the Be
stars. These are not "primordial" remnant accretion disks anymore, instead,
their origin must be found at the Main Sequence.  Their formation mechanism
is however not well understood, and an active community works on the
topic. The Be stars will be covered by many in these proceedings (see Stee for
a review on observations, Bjorkman and Jones on theory in these proceedings).

\section{The later stages, the post-Main Sequence}

The circumstellar material that is found in the post-Main Sequence
phase is due to mass loss from the star.  The circumstellar matter can
be found in many shapes, ranging from spherical shells to disky,
axi-symmetric, structures. The latter are most likely responsible for
the shaping of the Planetary Nebulae (PN), while rapidly rotating
stars are now predicted to be Gamma-Ray Burst progenitors and the mass
loss of such objects will be primarily concentrated in the equatorial
plane.

Entire conferences have been dedicated to the aspherical PNe
\citep{zijlstra_2011}. Whereas their progenitors, the AGB stars are mostly
round on larger scales, the PNe show a wide range of, mostly, axi-symmetric
morphologies ranging from nearly round elliptical nebulae to highly collimated
bipolar jet-like geometries.  The search is on for the origin for the change
in geometry as the stars evolve from the AGB to the post-AGB to the PN stage
\citep{vanhoof_1997}. This may have its foundations in the clumpy winds that
have been seen at very small scales in the AGB phase \citep{weigelt_2002}. The
subsequent phase in evolution is when the objects are post-AGB stars. These
objects have long been inferred to be surrounded by disks based on indirect
evidence such as their SEDs (e.g. \citealt{deruyter_2006}).  Direct evidence
for disks with inner radii of order 10 au has recently come from
interferometric measurements \citep{deroo_2007, lykou_2011}.  There seems to
be a strong correlation of the presence of bipolar PNe with binarity. The
objects which have direct evidence for disks in the papers mentioned above are
no exception, as they are also located in binary systems.

The situation for higher mass is more unclear. There is a diverse
selection of, often very variable, objects to be found in the upper
regions of the HR diagram, and even their evolutionary connections are
not always clear. A crude evolutionary scenario for objects around the
25-40$M_{\odot}$ mass can be summarized as the following

\begin{center}

 O star $\rightarrow$ (Red Supergiant) $\rightarrow$ Yellow
 Hypergiant/post-Red Supergiant $\rightarrow$ Luminous Blue Variable
 /B[e] $\rightarrow$ Wolf-Rayet $\rightarrow$ Supernova

\end{center}

Members of most of these evolutionary groups are plotted in the
HR-diagram in Figure~2. The various phases occupy complementary parts
in the diagram and define a more-or-less well defined sequence.

The Red Supergiants entry in the sequence above is put between
brackets, as these stars have not been observed and are not expected
at the highest masses.  The Yellow Hypergiants which are (obviously)
located bluewards of the RSG branch can be either pre- or post-RSG,
explaining why the extra qualification post-RSG is added in the
evolutionary sequence.  I also took the liberty to add the B[e]
supergiants, which, surprisingly perhaps, are not often explicitly
mentioned in evolutionary schemes such as this one. Their location in
the HR diagram is however well established for the Magellanic Cloud
sources \citep{zickgraf_1986}, and with more than 10 known, they are
as numerous as the LBVs in the Magellanic Clouds too.  If not the
direct descendants of the LBVs as Figure 2 might suggest, the B[e]
supergiants are conceivably in a similar stage of their evolution.

\articlefigure{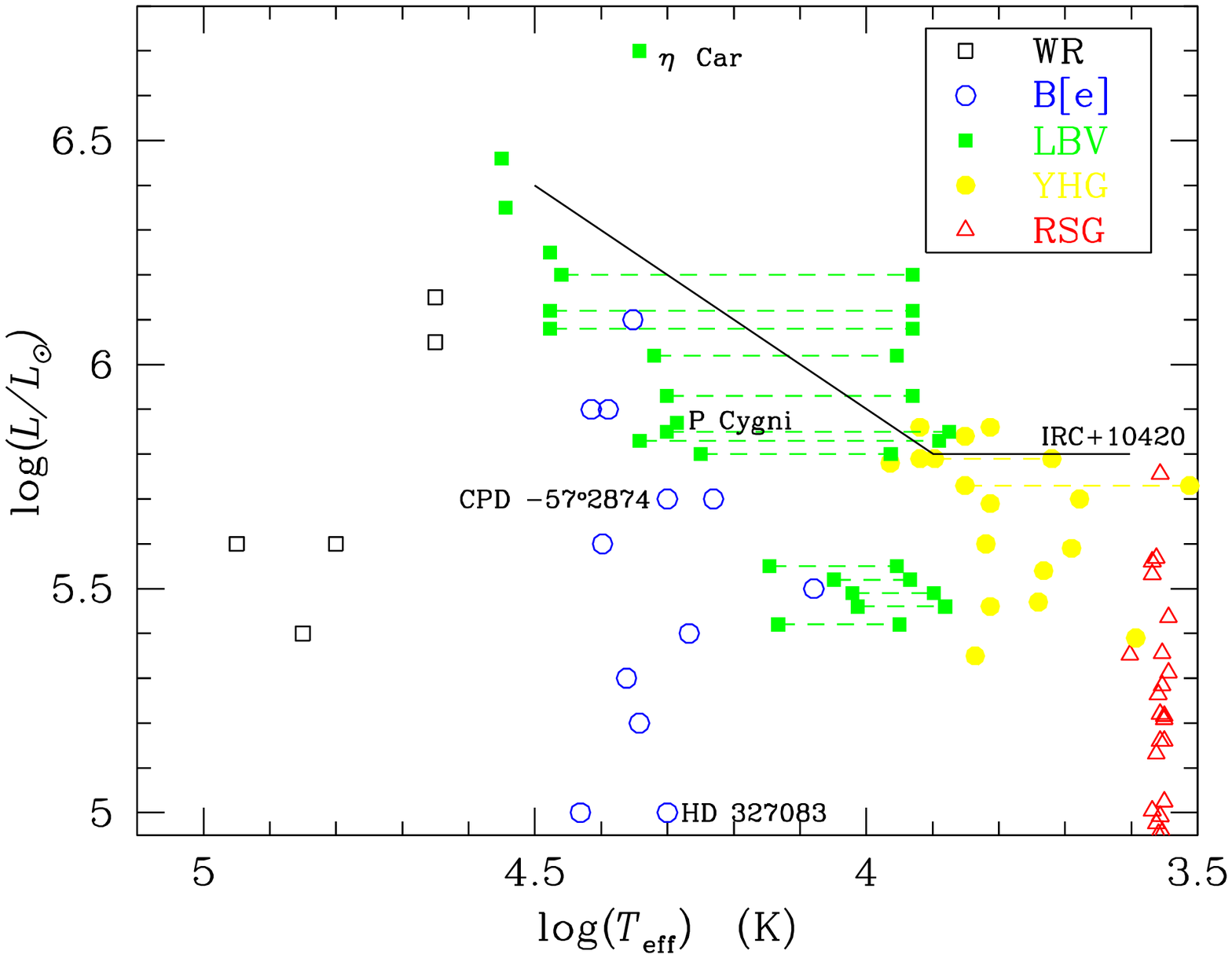}{lab1_fig} {Updated version of the HR-diagram
from \citet{rene_2009}, with some data taken from
\citet{smith_2004}. Shown are both Galactic and
Magellanic Cloud Luminous Blue Variables (LBV), Yellow Hypergiants
(YHG) and Red Supergiants (RSG). Added to the Galactic B[e] objects
mentioned in the text, are the Magellanic Cloud B[e] supergiants with
their data taken from \citet{zickgraf_2006}, the WN/WR stars mentioned
in \citet{vink_2011} whose parameters are taken from
\citet{hamann_2006}. Objects which are mentioned in the text are
indicated with their name. The solid line denotes the
Humphreys-Davidson limit.}

Let us discuss what we know about disks around the objects in their
evolutionary order.  A more general overview of the circumstellar
material around massive evolved stars can be found in
\citet{smith_2011}. Direct evidence for disks around O stars is
currently lacking, no interferometric imaging studies have yet
reported any. The Oe stars and variants on their spectral types, have
emission lines whose appearance suggest the presence of disks in the
same manner as for the Be stars. Having said that, spectropolarimetric
studies do not yield any detections of line-effect and thus do not
provide evidence for disks based on spectropolarimetry (see
\citealt{vink_2009} and references therein). However, we should bear
in mind that this absence of evidence for disks should not be taken as
evidence for the absence of the disks, so the question of disks around
O stars is still open. Moving to the Red Supergiants, despite some of
these being amongst the largest stars on the sky, disks have not been
observed. Diffraction limited (AO-assisted) imaging of objects such as
Betelgeuse and $\mu$ Cep reveal complex and clumpy environments
(\citealt{kervella_2011, dewit_2008a}). These could be the result of
mass loss which is possibly linked to the large spots on their stellar
surfaces. These stars are large enough that interferometric studies
have been mostly used to map the surfaces of these huge stars
\citep{ohnaka_2011}.

Direct information on the inner part of the circumstellar material
around the post-Red Supergiants is also sparse. The best known member
of the class, IRC +10420, has been the subject of several VLTI/AMBER
studies, \citep{dewit_2008, driebe_2009}. The lack of baselines in the
former did not allow an unambiguous determination of the geometry,
while the fact that a spherically symmetric radiative transfer model
did not fit the data, led the latter to conclude that the inner parts
are a-spherical. We recently obtained new data and compiled all data
available to re-visit the issue \citep{oud_dewit}. The number of
baselines and baseline coverage is arguably the best that can be
achieved for this equatorial target.  Of particular note is that the
differential phases across the Br$\gamma$ emission line can be best
explained by the fact that across the line, the second lobes of the
visibilities are probed, giving a characteristic flip in phase
angle. This allows a better handle on the geometries involved. The
best analytic model that can reproduce the data is a two ring model,
reminiscent of the rings observed towards SN1987A, with a small
circumstellar disk which blocks the light from the receding
ring. Ironically, although the optical interferometry discussed here
can provide direct evidence for disks, in this particular case, its
evidence is mostly indirect.

Moving on to the LBVs, Groh (these proceedings) discusses the innermost
regions of $\eta$ Car, and finds that the geometry can be best described as a
torus-like structure - as opposed to a disk which for the purposes of this
review can be best defined here as a structure whose scaleheight at a given
radius is (much) smaller than the radius. It is not obvious whether the binary
companion of $\eta$ Car has anything to do with the mass loss in the
equatorial regions however. When considering the other LBVs, no evidence for
disks has been published. Interferometry of the H$\alpha$ line of P Cygni (a
very rare type of observation), resulted in a round picture
\citep{balan_2010}.  Indirectly, most objects show signatures in their
spectropolarimetry similar to those in Be stars. Repeat observations showed
that the polarimetric signatures are not consistent with the disk hypothesis
however; the spectropolarimetric variability indicates the ejection of clumpy
material in random directions instead (\citealt{davies_2005}).

So, are any evolved stars associated with circumstellar disks?  The best
candidates would seem to be the B[e] supergiants.  The B[e] stars in the
Magellanic Clouds have long been suggested to be surrounded by disks based on
their hybrid spectral appearance. Broad emission and absorption lines can be
identified with a fast polar wind, while lower excitation and narrower lines
find their origin in a disk (\citealt{zickgraf_1985}), consistent with
spectropolarimetry (\citealt{magalhaes_2006}). The objects are thought to be
very rapid rotators, which may explain the preponderance of disks amongst
them. Unfortunately, the distance to the Magellanic Clouds is prohibitive for
detailed disk studies and we have to resort to the B[e] stars' galactic
cousins, which happen to constitute a diverse collection of objects. The
identification is particularly cumbersome as the distances, and thus
bolometric luminosities of the objects are hard to come by, hampering the
usual distinction between young and evolved stars.  In addition, it turns out
that a wide variety of objects can adhere to the primary classification
criterion of B[e] stars, namely the presence of forbidden emission lines
(\citealt{miro_2007}).  Several of the galactic B[e] supergiants have now been
observed in interferometry. \citet{domiciano_2007} describe the particularly
compelling case of CPD -57$^{\rm o}$2874, where both AMBER interferometry
tracing the Br$\gamma$ emission and hot dust continuum and MIDI interferometry
tracing warmer dust reveal a disk.  Other detailed studies show a different
picture, and seem to converge on the idea that a substantial number of the
accepted Galactic B[e] supergiants are members of binaries (HD 327803,
\citealt{wheelwright_2012}; HD 87643, \citealt{millour_2009}). It is
intriguing that the only massive evolved stars with evidence for disks are
often members of binary systems.

We finish this quick tour of the HR diagram with the Wolf-Rayet stars. They
are on average too faint yet to be systematically probed by
interferometry. Larger scale clumpy, round winds have been observed. Closer
in, the evidence for deviations from spherically symmetry is sparse. Indeed,
although no disks have been directly observed towards the WR-stars at present,
indirect evidence for disk-like structures has been used to strengthen their
status as GRB progenitors.  \citet{harries_1998} find a minority of WR stars
to exhibit spectropolarimetric signatures indicating deviations from spherical
symmetry.  \citet{vink_2011} used this fact to hypothesize that these objects
can be the elusive rapidly rotating stars that are suggested to be the
GRB-progenitors. Normally, rotation rates of WR stars are difficult to measure
because of their strong line emission, but an equatorial wind can be best
explained by rapid rotation.  They went further and note the fact that this
WR-subsample are associated with ejection nebulae from the previous RSG/LBV
phase (as found by comparing with the sample of \citealt{stock_2010}). Older WR
stars will have spun down and their ejection nebulae would be long dispersed
in to the interstellar medium, and be invisible. In other words, the objects
presented by \citet{vink_2011} are young, rapidly rotating, WR stars, and
among the strongest GRB-progenitor candidates.

To conclude this section. It would seem that disks such as clearly observed
around young stars are not present in the same abundance or as clear in the
post-Main Sequence phase of evolution. Clumpy material and binarity play a
large role, and often flattened structures are associated with the
binaries. The objects with the strongest - yet still indirect - evidence for
disks, the B[e] supergiants are thought to be rapid rotators, and when linking
this with the current thoughts on GRBs, can be a serious contender for their
progenitors.

\section{Final Word}

This overview presented some of the latest results in high resolution
studies of disks in various evolutionary phases. By the turn of the
century, we had to rely mostly on indirect evidence for the presence
of small scale disks around objects. The most compelling observations
at that time were limited to the bright and nearby Be stars. As a
result these objects acted as the pathfinder, not only then, but also
now in ever more refined disk studies.  We found that with the
advances in optical/infrared infrared interferometry,
spectro-astrometry, IFU spectroscopy often combined with routine AO
support, {\it direct} evidence for disks has come forward in abundance
this century. The next steps will be to put the samples on a more
statistical footing, and derive astrophysical parameters for them to
finally and fully understand what is happening. The leaps in both
observational techniques and theoretical and numerical models to
interpret those new data over the past decade are remarkable and
promise an exciting future.

\acknowledgements RDO wishes to thank Alex Carciofi and Thomas
Rivinius for organizing a very interesting conference in an equally
interesting location. He also is grateful to Ben Davies, Willem-Jan
de Wit, Janet Drew, Jorick Vink and Hugh Wheelwright on our journey
towards ever high resolution and thanks WJ, Hugh and Jorick also for
carefully reading this manuscipt.

\bibliography{csdyn_oud}

\end{document}